\documentclass{article}

\usepackage{graphicx}
\usepackage[round]{natbib}
\usepackage{epsfig}

\title{
Improving the expected accuracy of forecasts of future climate using a simple bias-variance tradeoff
}
\author{
Stephen Jewson (RMS) \footnote{\emph{Correspondence email}: \texttt{stephen.jewson@rms.com}} \\
and \\
Ed Hawkins (NCAS-Climate, University of Reading)
}

\begin{document}
\maketitle

\begin{abstract}
We describe a simple method that utilises the standard idea of bias-variance trade-off to improve the
expected accuracy of numerical model forecasts of future climate.
The method can be thought of as an optimal multi-model combination between the forecast from a
numerical model multi-model ensemble, on one hand, and a simple statistical forecast, on the other.
We apply the method to predictions for UK temperature and precipitation for the period 2010 to 2100.
The temperature predictions hardly change, while the precipitation predictions show large changes.
\end{abstract}

\section{Introduction}

Consider the following two (entirely fictitious) climate forecasts:

\begin{description}
  \item[(a)] the expected annual mean temperature in 2050 will be 2 degrees warmer than now. The range of annual mean temperatures will be the same as now. The standard error on the predicted change is 0.1 degrees.

  \item[(b)] the expected annual mean temperature in 2050 will be 2 degrees warmer than now. The range of annual mean temperatures will be the same as now. The standard error on the predicted change is 2.0 degrees.
\end{description}

The forecasts for the expected temperature are the same, and the forecasts for the level of variability are the same (and are both predicting the same variability as now).
The only difference is in the uncertainty on the change in the mean, as represented by the standard error.
On intuitive grounds, this should lead to a very different response from someone who is concerned about annual mean temperature in 2050. Given the first forecast (if they believe it, including believing the estimate of uncertainty) then they might well take action to deal with the possible consequences of the predicted change. Given the second forecast (again, if they believe it) then they might well not take action, since the uncertainty is so great. Instead they might wait until closer to the time, and hope for a less uncertain forecast to come along, or they might just decide to prepare themselves for a range of possibilities. Clearly the details of that decision will be very dependent on the individual, but the principle is clear: the greater the uncertainty, the less useful the forecast.

Now consider a third example:

\begin{description}

  \item[(c)] the expected annual mean temperature in 2050 will be 1 degree warmer than now. The range of annual mean temperatures will be the same as now. The standard error on the predicted change is 0.1 degrees.

\end{description}

The predicted change in the expected temperature is smaller than in (a) and (b),
while the uncertainty is the same as in (a).
In this case, one could imagine that the decision maker might decide to take action now.
From this we can see intuitively that it is not the size of the forecasted change which is likely to lead to action now, since the forecasted change in (b) is larger than the forecasted change in (c).
Rather, the decision to take action would seem to be somehow related to the ratio between the size of the forecasted change and the uncertainty around that change {i.e.} a type of signal-to-noise ratio. If the change is large relative to the size of the uncertainty, then it seems likely it might make more sense to take action now.
If the change is small relative to the size of the uncertainty, then it seems more reasonable to wait and see.

There are a number of ways that one might try to put these ideas into a mathematical framework.
For instance, it is sometimes argued that one should fold all sources of variability and uncertainty into a single distribution, and then use utility theory to make decisions based on that distribution (for example~\citet{bernardo} follow this line of reasoning).
Another possibility is to present the data as is: the predicted expected value, the predicted variability, and the estimated uncertainty, and leave it at that.
In this article, however, we describe a very simple method for \emph{adjusting} a forecast, based on the size of the uncertainty.
The three main motivations for the method can be summarized as follows:
\begin{description}
  \item [(a)]
The method is based on a mathematical version of the idea described above that it is the ratio of the prediction to the size of the uncertainty that really matters, not the prediction itself.
This ratio (in some form: see below) is then used to adjust the forecast. Essentially the method converts a forecast of a large change, but with large uncertainty, into a smaller forecast with lower uncertainty

  \item [(b)]
The method is based on the idea of trying to make the
prediction is as accurate as possible (in some specific sense of the word accurate, as explained below).

 \item [(c)]
When climate forecasts from numericals models are very uncertain it seems to make intuitive sense to start mixing the climate model forecasts with simple statistical forecasts, in some kind of optimal multi-model ensemble.

\end{description}

We call the method `damping', and the resulting forecasts `damped forecasts'. However, similar methods have gone by other names in other situations. Statisticians sometimes use the phrase `shrinkage' to describe something similar, although they also use the term shrinkage more generally too. The method can be considered as an example of the standard mathematical modelling concepts known as regression-towards-the-mean, regularization and bias-variance tradeoff.

\section{Damped forecasts}

The set-up we use to derive our damped forecasts is as follows.

Let $y$ be the annual mean temperature, measured at London Heathrow (LHR) in 2050.
Note that $y$ is a measurable quantity, unlike the \emph{expected} annual mean temperature in 2050.

How might we predict $y$? We will assume we have two sources of information:
historical station data, and a numerical climate model prediction.
The climate model prediction could be an initial condition ensemble, or a multi-model ensemble,
but needs to include an estimate of the uncertainty.

Given these two sources of information,
one option would be to take a prediction directly from the climate model, and ignore the station data.
However, climate models do not typically represent the range of scales of variability necessary to
capture real measured values, and so a more usual (and likely more accurate) approach would be
to use the climate model not to predict absolute values, but rather just the change. The assumption behind
this is that the change in temperature at LHR is going to be highly correlated with large-scale changes, which the climate
model is designed to capture. The absolute level of temperature at LHR can then be taken from observations.

Given this (which, so far, is standard reasoning) we can say:
\begin{equation}
y=c+d+e
\end{equation}
where
\begin{itemize}
  \item $c$ is the expected temperature for 2008 (which is unknown)
  \item $d$ is the deterministic change in the expected temperature between 2008 and 2050, driven by CO$_2$ and other forcings
  \item $e$ is the internal variability contribution to 2050 annual mean temperature
\end{itemize}

The most obvious way to predict $y$ is then:
\begin{equation}\label{eq1}
\hat{y}=\hat{c}+\hat{d}
\end{equation}

In other words: we try and estimate the expected temperature for 2008, and we add on an estimate for the
change.

We will take $\hat{c}$ from station data (maybe using an average of the last 10 years) and $\hat{d}$ from the climate model (most likely as an average of an ensemble, ideally over initial conditions, parameters and models). Both estimates are uncertain.

As usual in prediction problems, the prediction $\hat{y}$ can be considered to be a prediction of the actual value of $y$, or a prediction of the expected value of $y$: we will consider it to be a prediction of the actual value of $y$.

Note that we could also make a probabilistic prediction for $y$. If we assume that $y$ is normally distributed, then we could write this as:

\begin{equation}
\hat{y} \sim N(\hat{c}+\hat{d},\hat{\sigma}^2)
\end{equation}

where we have included an estimate for the variance.

The climate model prediction contains uncertainty, for a variety of reasons including
model error, scenario error and sampling error from having a finite ensemble. As discussed in the introduction,
if that uncertainty is extremely large, then it seems intuitively reasonable that we might actually be better off ignoring the model completely,
and predicting $y$ using the simple prediction:
\begin{equation}\label{eq1.5}
\hat{y}=\hat{c}
\end{equation}
On the other hand, if the uncertainty is small then we would probably be happy with the prediction from equation~\ref{eq1}.

Our damped prediction is a unification of these two limiting cases, and is given by:

\begin{equation}\label{eq2}
\hat{y}=\hat{c}+k\hat{d}
\end{equation}

where $k$ is between $0$ and $1$, and is determined by the level of uncertainty in the model prediction.
For a highly uncertain prediction, $k$ should be close to, or equal to, zero, and the climate model should
be ignored. For a prediction with low uncertainty, $k$ should be close to one, and the prediction from the
climate model should be embraced. The value of $k$ would be expected to vary in space and time, and from variable to variable, following the variation in the uncertainty in forecasts.

Again we can make a probabilistic forecast, which now becomes:
\begin{equation}\label{eqprob}
\hat{y} \sim N(\hat{c}+k \hat{d},\hat{\sigma}^2)
\end{equation}

The variance would now be different: expressions for the appropriate variance are derived in the appendix.

We note that equation~\ref{eq2} is not symmetric in $\hat{c}$ and $\hat{d}$: this is because $\hat{c}$ is a prediction of
the \emph{absolute temperature}, while $\hat{d}$ is a prediction of a \emph{change} in temperature. While it does make sense to
`damp' the prediction back towards $\hat{c}$ using $k$, it would not make sense to `damp' $\hat{c}$ towards $\hat{d}$.

Equation~\ref{eq2} represents a combination between the climate model forecast of change $\hat{d}$ with a `simple model' forecast
of no change at all. An alternative would be to consider combining the climate model forecast with a slightly more complex
simpler model, that is still independent of the climate model. For temperature, this might be a linear trend. This is discussed in more detail
in the discussion in section 6.

In the next two sections we derive quantitative expressions for $k$, first based on the idea of
minimizing predictive MSE, and secondly based on the idea of minimizing predictive likelihood.

\section{Deriving the damping using predictive MSE}

We now derive an expression for $k$ using the idea that $k$ should be chosen to minimise the mean square error (MSE) of the forecast.
The MSE for the forecast given by~\ref{eq2} is:

\begin{eqnarray}\label{eq3}
\mbox{MSE}
&=&E[(\mbox{error})^2]\\
&=&E[(y-\hat{y})^2]\\
&=&E[(c+d+e-\hat{c}-k\hat{d})^2]\\
&=&E[((c-\hat{c})+(d-k\hat{d})+e)^2]\\
&=&E[(c-\hat{c})^2+(d-k\hat{d})^2+e^2+\mbox{cross terms}]\\
&=&E[(c-\hat{c})^2]+E[(d-k\hat{d})^2]+E[e^2]+0\\
&=&\mbox{MSE}_c+\mbox{MSE}_d+E[e^2]
\end{eqnarray}

The cross-terms disappear because we assume that the variability around the expected value in 2050 will be
uncorrelated with the errors in our estimation of $c$ and $d$, and because the errors in our estimation of
$c$ and $d$ are uncorrelated with each other. These both seem to be reasonable assumptions.

We see that the total MSE breaks down into a term due to the MSE in the prediction of the expected temperature for 2008,
a term due to the MSE in the prediction of the change in temperature between 2008 and 2050, and a term due to
variability in temperature in 2050.

The MSE$_c$ term breaks down further:

\begin{eqnarray}
\mbox{MSE}_c
&=&E[(c-\hat{c})^2]\\
&=&E[c^2-2c\hat{c}+\hat{c}^2]\\
&=&E[c^2]-2E[c\hat{c}]+E[\hat{c}^2]\\
&=&c^2-2cE[\hat{c}]+E[\hat{c}^2]\\
&=&c^2-2c^2+E[\hat{c}^2]\\
&=&E[\hat{c}^2]-c^2\\
&=&V[\hat{c}]
\end{eqnarray}

where we have assumed that our estimate of $c$ is unbiased, which seems reasonable.

We see that MSE$_c$ is equal to the uncertainty (aka the standard error) on the estimate of $c$, and does not depend on $k$.
If we calculate $\hat{c}$ using, say, a 10 year mean or a 30 year linear trend, then it is easy enough to estimate this variance using standard statistical formulae.

The MSE$_d$ term can be decomposed as follows:
\begin{eqnarray}
\mbox{MSE}_d
&=&E[(d-k\hat{d})^2]\\
&=&E[d^2-2dk\hat{d}+k^2\hat{d}^2]\\
&=&E[d^2]-2E[dk\hat{d}]+E[k^2\hat{d}^2]\\
&=&d^2-2dkE[\hat{d}]+k^2E[\hat{d}^2]\\
&=&d^2-2d^2k+k^2(V(\hat{d})+d^2)\\
&=&d^2(1-2k+k^2)+k^2V(\hat{d})\\
&=&d^2(k-1)^2+k^2V(\hat{d})
\end{eqnarray}

where we have assumed that our estimate of $d$ is unbiased too, which also seems reasonable.

The first term in this decomposition is the effect of the bias introduced by the damping (it is actually the bias squared).
If $k=1$ (the undamped prediction) then the bias is zero. If $k=0$ (pure observed data, no climate model data) then the bias effect is $d^2$.

The second term is the model error variance, that captures the fact that model uncertainty increases the MSE.
This is the term that we seek to decrease using damping.

We now write
\begin{equation}
E[e^2]=\sigma^2
\end{equation}
This is the internal variability contribution to the MSE.

Recombining all the terms in equation~\ref{eq3}, we now have
\begin{eqnarray}\label{eqmse}
\mbox{MSE}=V[\hat{c}]+d^2(k-1)^2+k^2V(\hat{d})+\sigma^2
\end{eqnarray}

All the dependence on $k$ has now been made explicit in this equation, and so we can think about
varying $k$ to minimise the MSE.
Our damping method tries to find the value of $k$ that balances the bias and the variance terms
in this equation. Values of $k$ near to 1 give a small bias but a large variance, and hence a large MSE.
Values of $k$ near to 0 give a large bias but a small variance, and hence again a large MSE.
Values of $k$ between the extremes give the lowest MSE.

Since equation~\ref{eqmse} is quadratic in $k$, minimising it is easy. If we differentiate wrt $k$ we get:
\begin{eqnarray}
\frac{\partial}{\partial k}\mbox{MSE}=0+2 d^2(k-1)+2kV(\hat{d})+0
\end{eqnarray}

Setting this equal to zero gives:
\begin{eqnarray}\label{eq4}
k=\frac{d^2}{d^2+V(\hat{d})}
\end{eqnarray}

This gives the value of $k$ that minimises MSE.
Consideration of the form of this expression shows that it behaves exactly as expected intuitively.
For instance:
\begin{itemize}
  \item when the uncertainty on the predicted change is very large relative to the change itself, then
  $k$ is very small and the predicted change will be more or less ignored
  \item when the uncertainty on the change is comparable to the change itself, $k$ is around 0.5,
  and the forecasted change will be reduced to around half the original value
  \item when the uncertainty on the change is small compared to the change itself then $k$ is close to 1
  and the forecasted change is believed in its entirety
\end{itemize}

\section{Deriving the damping using predictive likelihood}

We now repeat the exercise of deriving the optimal damping, but thinking probabilistically.
In other words, we choose $k$ to maximise predictive skill as measured using a measure of the skill
of the probabilistic forecast that we will produce, rather than the point forecast.
The probabilistic measure we choose is the expected log likelihood.

The expected log-likelihood of the probabilistic forecast given by equation~\ref{eqprob} is:
\begin{eqnarray}
c
&=&E[\mbox{log}p(y,\hat{\mu},\hat{\sigma},k)]\\
&=&E[\mbox{log}N(y,\hat{\mu}=\hat{c}+k\hat{d},\hat{\sigma})]\\
&=&E\left[\mbox{log}\left(\frac{1}{\sqrt{2\pi}}\frac{1}{\hat{\sigma}}\mbox{exp}\left(-\frac{(y-\hat{\mu})^2}{\hat{\sigma}^2}\right)\right)\right]\\
&=&E\left[ -\mbox{log}\sqrt{2\pi} -\mbox{log} \hat{\sigma} -\frac{(y-\hat{\mu})^2}{2\hat{\sigma}^2}\right]\\
&=&E\left[ -\mbox{log}\sqrt{2\pi} -\mbox{log} \hat{\sigma} -\frac{(y-\hat{c}-k\hat{d})^2}{2\hat{\sigma}^2}\right]
\end{eqnarray}

Differentiating wrt $k$ gives:
\begin{eqnarray}
\frac{\partial c}{\partial k}
&=&0+0-\frac{\partial}{\partial k} E\left[\frac{(y-\hat{c}-k\hat{d})^2}{2\hat{\sigma}^2}\right]\\
&=&E\left[\frac{(y-\hat{c}-k\hat{d})\hat{d}}{\hat{\sigma}^2}\right]\\
&=&E\left[\frac{y\hat{d}-\hat{c}\hat{d}-k\hat{d}^2}{\hat{\sigma}^2}\right]\\
&=&E\left[\frac{y\hat{d}}{{\hat{\sigma}^2}}\right]
  -E\left[\frac{\hat{c}\hat{d}}{{\hat{\sigma}^2}}\right]
  -E\left[\frac{k\hat{d}^2}{{\hat{\sigma}^2}}\right]\\
&=&E\left[y\right]E\left[\hat{d}\right]E\left[\frac{1}{\hat{\sigma}^2}\right]
  -E\left[\hat{c}\right]E\left[\hat{d}\right]E\left[\frac{1}{\hat{\sigma}^2}\right]
  -kE\left[\hat{d}^2\right]E\left[\frac{1}{\hat{\sigma}^2}\right]\\
&=&\left(E\left[y\right]E\left[\hat{d}\right]
        -E\left[\hat{c}\right]E\left[\hat{d}\right]
        -kE\left[\hat{d}^2\right]\right)
        E\left[\frac{1}{\hat{\sigma}^2}\right]\\
&=&\left((c+d)d-cd-k(V(\hat{d})+d^2)\right)E\left[\frac{1}{\hat{\sigma}^2}\right]\\
&=&\left(d^2-kV(\hat{d})-kd^2\right)E\left[\frac{1}{\hat{\sigma}^2}\right]
\end{eqnarray}

Setting equal to zero gives:
\begin{eqnarray}
k=\frac{d^2}{d^2+V(\hat{d})}
\end{eqnarray}

We see that it does not make any difference when we consider a probabilistic score: the optimal damping factor
is the same as for the point forecast case. This is because the expected log likelihood is made up of an MSE term,
and a spread term. The spread term depends only on the predicted spread, and is unaffected by the damping, and
so the problem reduces to minimising the MSE.

\section{Examples}

We now apply the expression for $k$ derived above to some real examples.
Since in equation~\ref{eq4} $k$ depends on two unknowns ($d$ and $V(\hat{d})$) this expression cannot be evaluated
directly. Instead, we need to come up with an estimator for $k$ that can be evaluated.
The most obvious choice for $d$ is the `plug-in' estimator, in which we use $\hat{d}$ for $d$.
For $V(\hat{d})$ we can use an estimate of the uncertainty around the predicted change based, in some
way, on the ensemble spread. The use of estimators for the terms in expression~\ref{eq4}, rather than the real
values, means that the optimality properties will not quite hold anymore: this is a standard problem that runs
through much of classical statistics. The hope is that the deterioration in the optimality
properties is not so great as to negate the benefit of the method as a whole. One would expect that
if the value of $k$ comes out very close to zero or one, then it is likely that the damping process brings no benefit,
and may even harm the forecast (relative to setting $k$ to exactly zero or one), while if the value of $k$ is
around 0.5, it is more likely that the damping process brings some benefit.

\subsection{Temperature example}

Our first example is based on the CMIP3 annual mean temperature predictions described in~\citet{cmip}.
The specific data we use is exactly the same data used in~\citet{hawkins09a}, for the A1B scenario, for the UK.
The estimates of the uncertainty ($V(\hat{d})$) that we use are derived from the subsequent analysis
of that data described in~\citet{jewson09a}. In that analysis, scenarios for the possible correlation between
the different models in the CMIP3 ensembles were converted into adjustments to the ensemble spread.
We assume that these adjustments include all sources of uncertainty, including all model and parameter uncertainty.
Figure 1 in~\citet{jewson09a} shows the original ensemble spread, and four scenarios based on
assumptions of correlations of 0.0, 0.25, 0.5 and 0.75 between the models.

Given these estimates of `corrected' ensemble spread (which can be thought of as what the ensemble spread
would be if the climate models were genuinely independent, and included all sources of uncertainty) we can derive $V(\hat{d})$ using
the standard $\sigma^2/n$ expression for the standard error on the mean, where $\sigma^2$ is the corrected
variance and $n$ is the number of models (which is 15).

\begin{figure}[!ht]\begin{center}
\scalebox{0.8}{\includegraphics{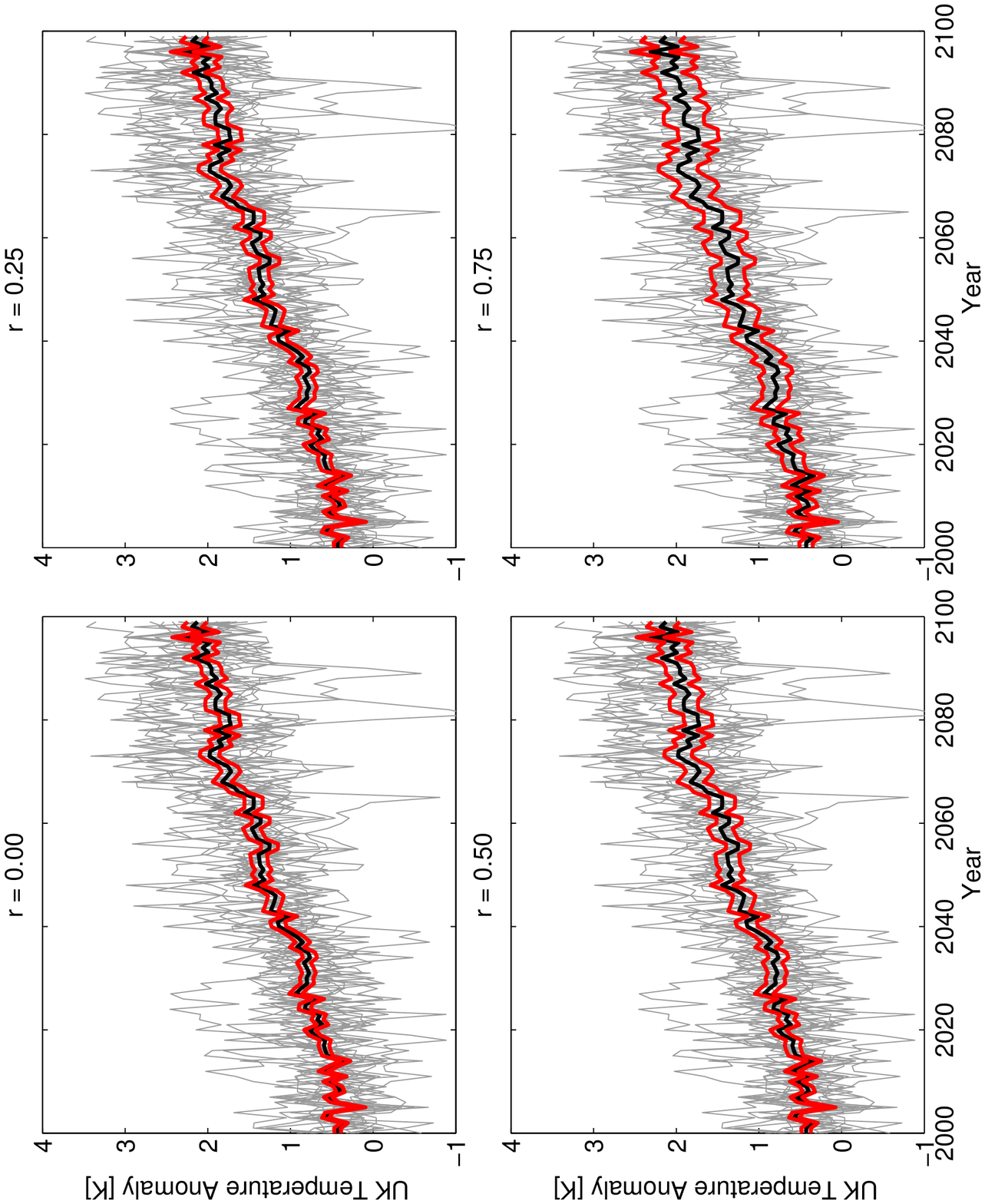}}
\end{center}\caption{
The estimated change in UK temperature versus time, from the CMIP3 ensemble (grey lines)
along with the ensemble mean (black line) and the estimated standard error on the mean
(shown as plus and minus one standard deviation).
The estimates of the uncertainty are based on correlation scenarios of:
(a) 0,
(b) 0.25,
(c) 0.5, and
(d) 0.75.
}
\label{fig1}\end{figure}

The four panels of figure~\ref{fig1} each show the members of the CMIP3 ensemble (grey lines, which are the same in all panels), 
the ensemble mean estimate for $d$ (black line, the same in all panels),
and the standard error on the ensemble mean (red lines, plus or minus one standard deviation).
The four panels show the scenarios of correlation equal to 0, 0.25, 0.5 and 0.75.
We see, as expected, that the standard error is largest furthest into the future, and for the case in which we assume
that the model results are the most correlated.

The four panels of figure~\ref{fig2} then show the estimates of the optimal $k$ derived from the data in figure~\ref{fig1}.
Up to around 2015, the values of $k$ fluctuate wildly.
Considering figure~\ref{fig1} we can see that this is because $d$ is close to zero, and is poorly estimated in the sense
that the estimate shows very large fractional fluctuations.
This, in turn, leads to large variations in the estimate for $k$.  These variations are unlikely to be meaningful, and it might
make sense to smooth them out, although we don't do that in the analysis described here.
We see values of $k$ between 0.8 and 1.0 for the $r=0$ correlation scenario,
dropping to values between 0.5 and 0.8 for the $r=0.75$ correlation scenario. For all scenarios the values of $k$ at longer
lead times are higher, since the predicted signal grows faster than the standard error.

Figure~\ref{fig3} shows the ensemble members and the ensemble mean as in figure~\ref{fig1}, 
but now with the damped ensemble mean forecast included as a blue line.
Overall the damping does not show a large impact, even for the $k=0.75$ scenario. 
The biggest fractional impact is in the first two decades of the predictions.

Figure~\ref{fig4} shows the estimated sizes of errors from making forecasts using `no change' (the black line),
the ensemble mean (the red line), and the damped ensemble mean (the blue line). 
The ensemble mean and damped ensemble mean forecasts always beat the no change forecast (apart from one or two spikes in the data
that are presumably just noise).
There is a small benefit from using the damped ensemble mean rather than the ensemble mean in the first two decades, but one might
doubt whether that benefit would be realised in practice because of the use of plug-in estimators.

\subsection{Precipitation example}

\begin{figure}[!ht]\begin{center}
\scalebox{0.8}{\includegraphics{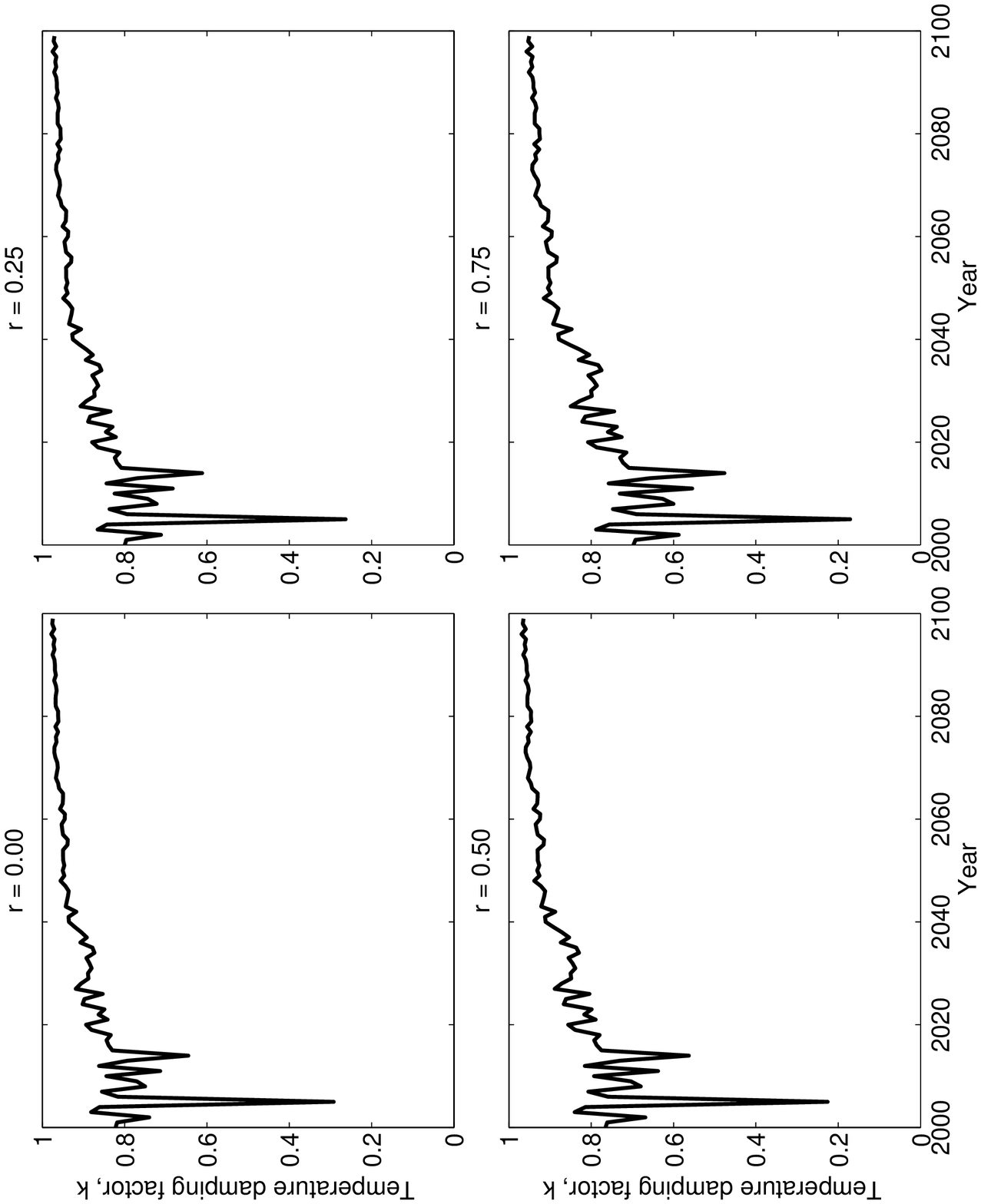}}
\end{center}\caption{
Estimates for the optimal damping factor $k$ derived from the data in figure~\ref{fig1}.
}
\label{fig2}\end{figure}

\begin{figure}[!ht]\begin{center}
\scalebox{0.8}{\includegraphics{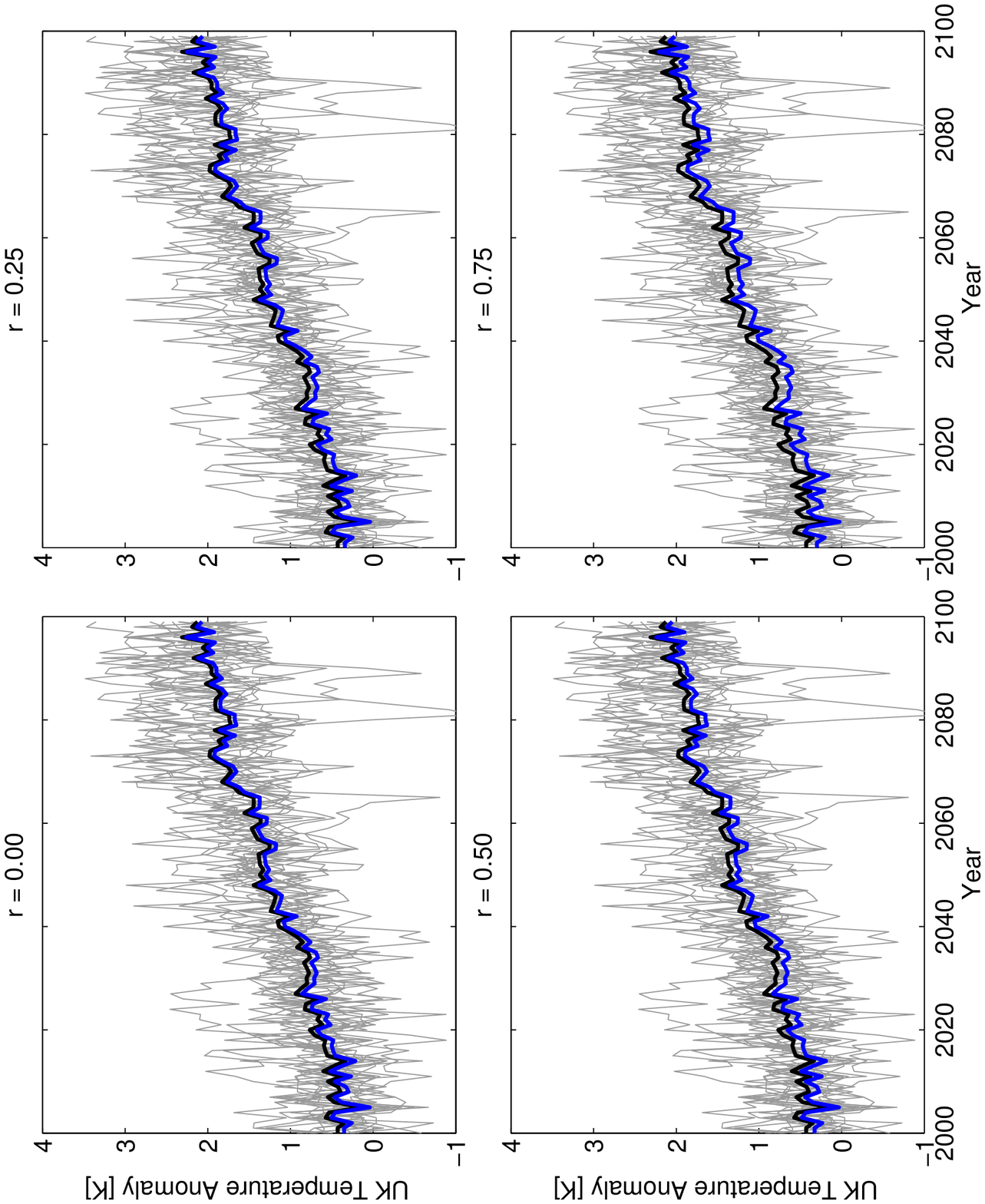}}
\end{center}\caption{
The estimated change in UK temperature versus time, from the CMIP3 ensemble (grey lines)
along with the ensemble mean (black line) and a damped version of the ensemble mean
(blue line).
The estimates of the uncertainty are based on correlation scenarios of:
(a) 0,
(b) 0.25,
(c) 0.5, and
(d) 0.75.
}
\label{fig3}\end{figure}

\begin{figure}[!ht]\begin{center}
\scalebox{0.8}{\includegraphics{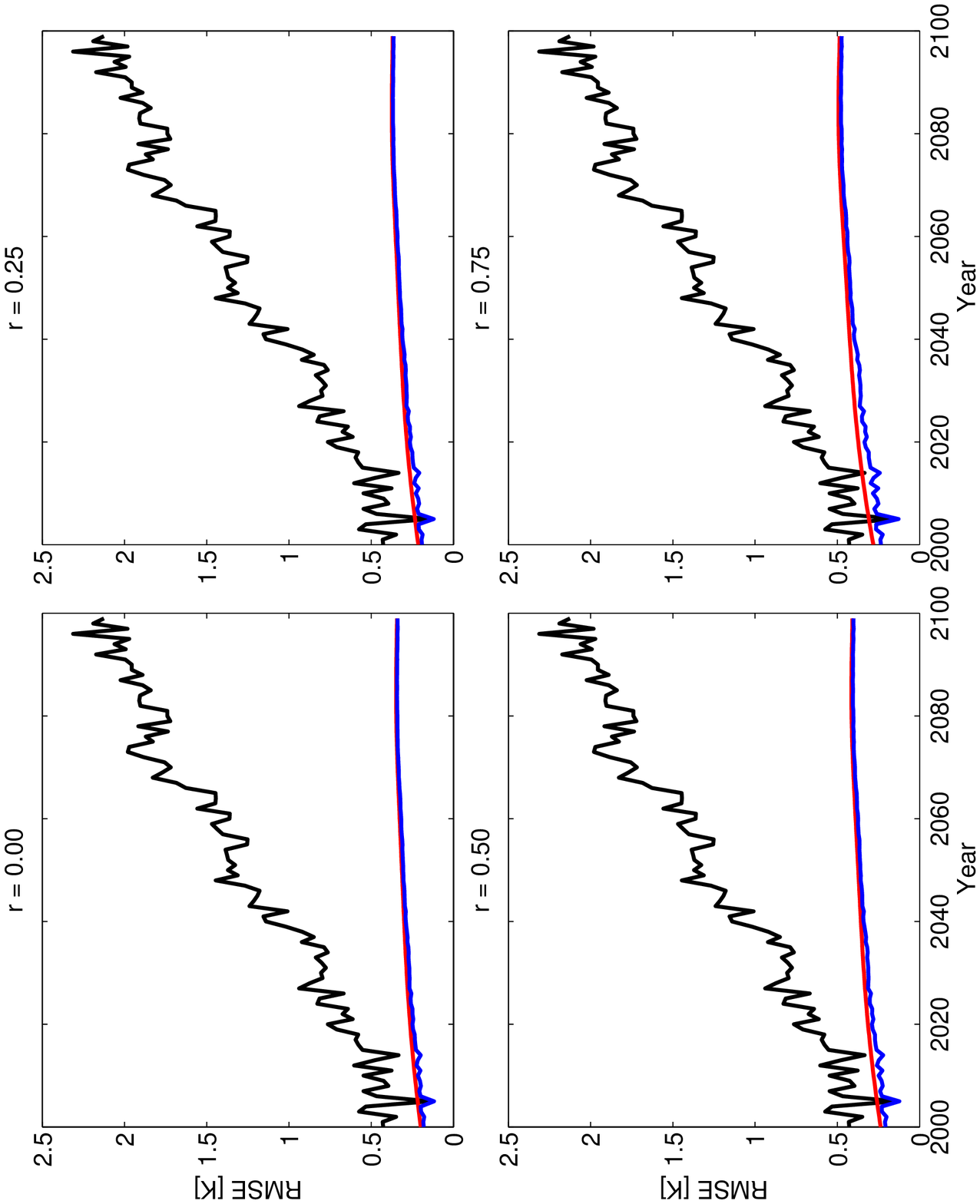}}
\end{center}\caption{
The estimated RMSE of temperature predictions made from the `no change' forecast (black line), the IPCC
ensemble mean (red line) and the damped IPCC ensemble mean (blue line).
}
\label{fig4}\end{figure}

\begin{figure}[!ht]\begin{center}
\scalebox{0.8}{\includegraphics{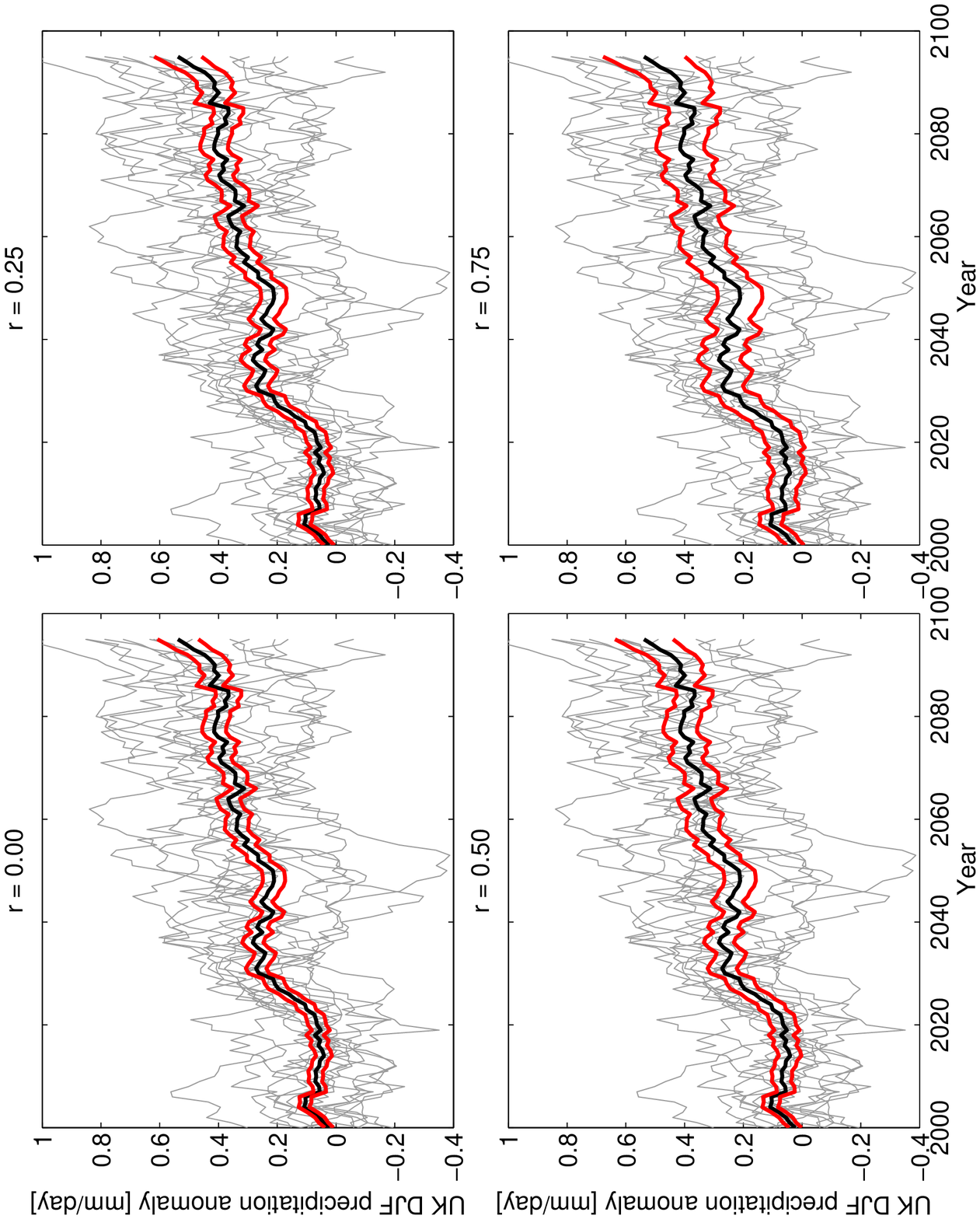}}
\end{center}\caption{
The estimated change in UK winter precipitation versus time, from the CMIP3 ensemble (grey lines)
along with the ensemble mean (black line) and the estimated standard error on the mean
(shown as plus and minus one standard deviation).
The estimates of the uncertainty are based on correlation scenarios of:
(a) 0,
(b) 0.25,
(c) 0.5, and
(d) 0.75.
}
\label{fig5}\end{figure}

\begin{figure}[!ht]\begin{center}
\scalebox{0.8}{\includegraphics{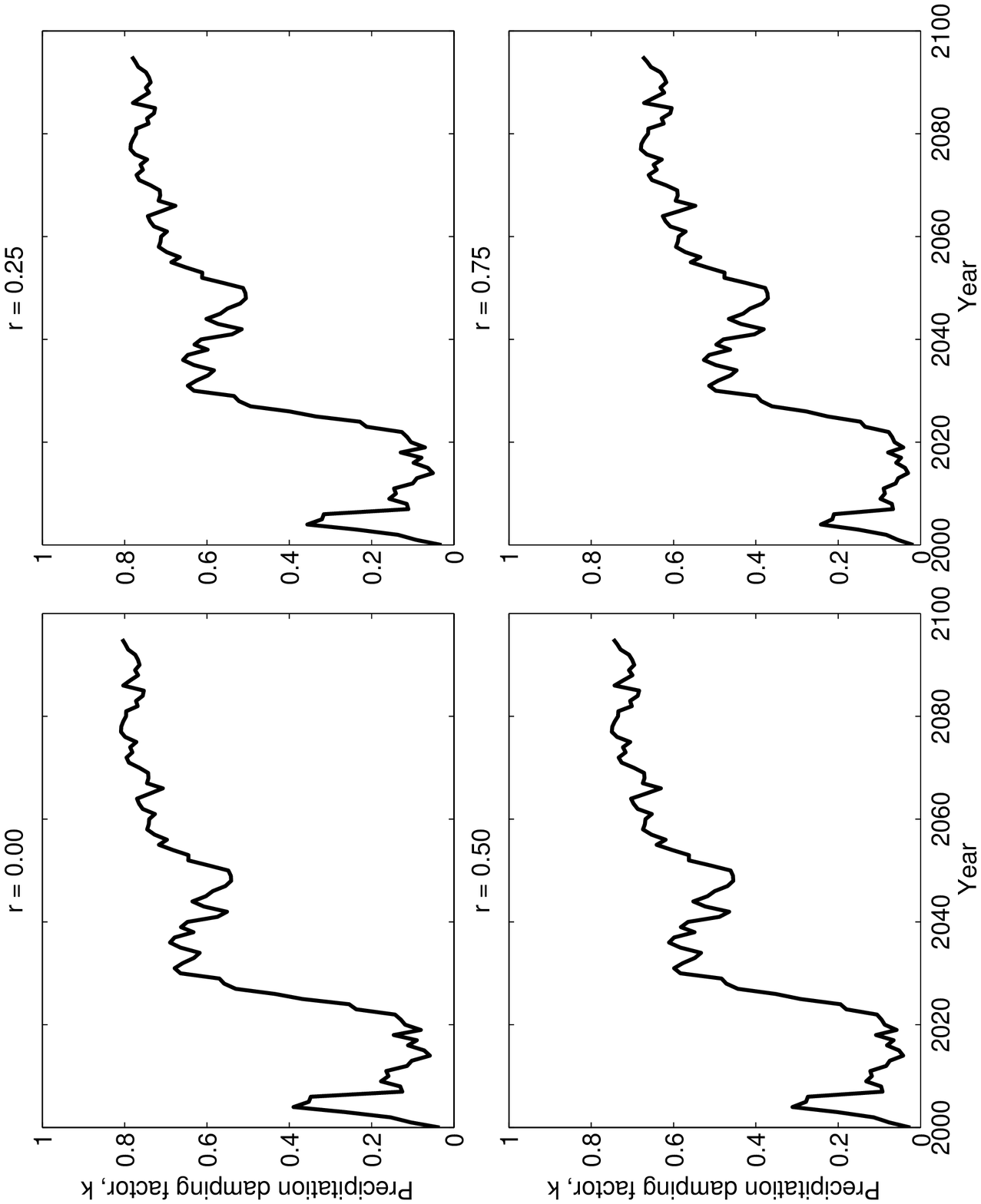}}
\end{center}\caption{
Estimates for the optimal damping factor $k$ derived from the data in figure~\ref{fig1}.
}
\label{fig6}\end{figure}

\begin{figure}[!ht]\begin{center}
\scalebox{0.8}{\includegraphics{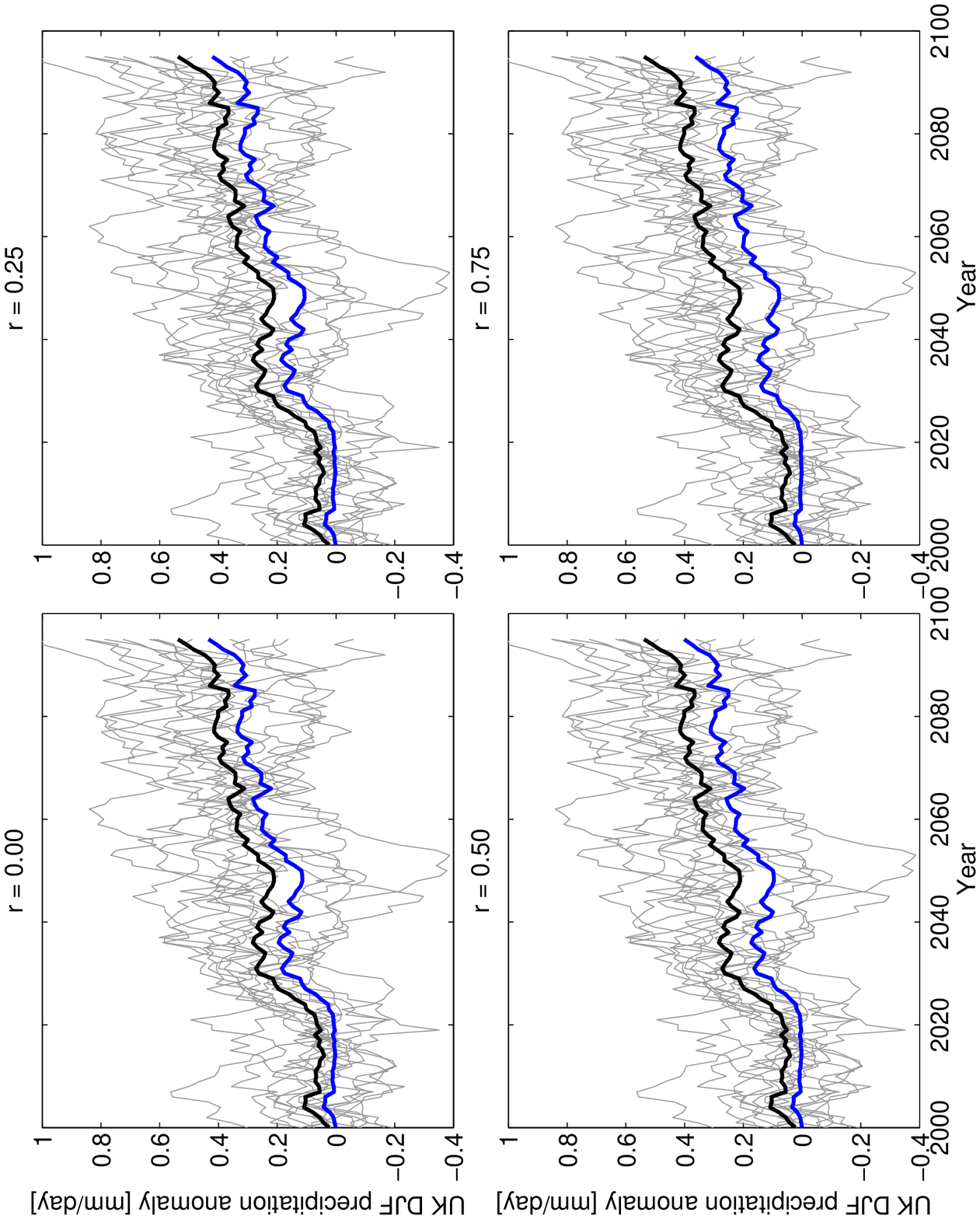}}
\end{center}\caption{
The estimated change in UK precipitation versus time, from the CMIP3 ensemble (grey lines)
along with the ensemble mean (black line) and a damped version of the ensemble mean
(blue line).
The estimates of the uncertainty are based on correlation scenarios of:
(a) 0,
(b) 0.25,
(c) 0.5, and
(d) 0.75.
}
\label{fig7}\end{figure}

\begin{figure}[!ht]\begin{center}
\scalebox{0.8}{\includegraphics{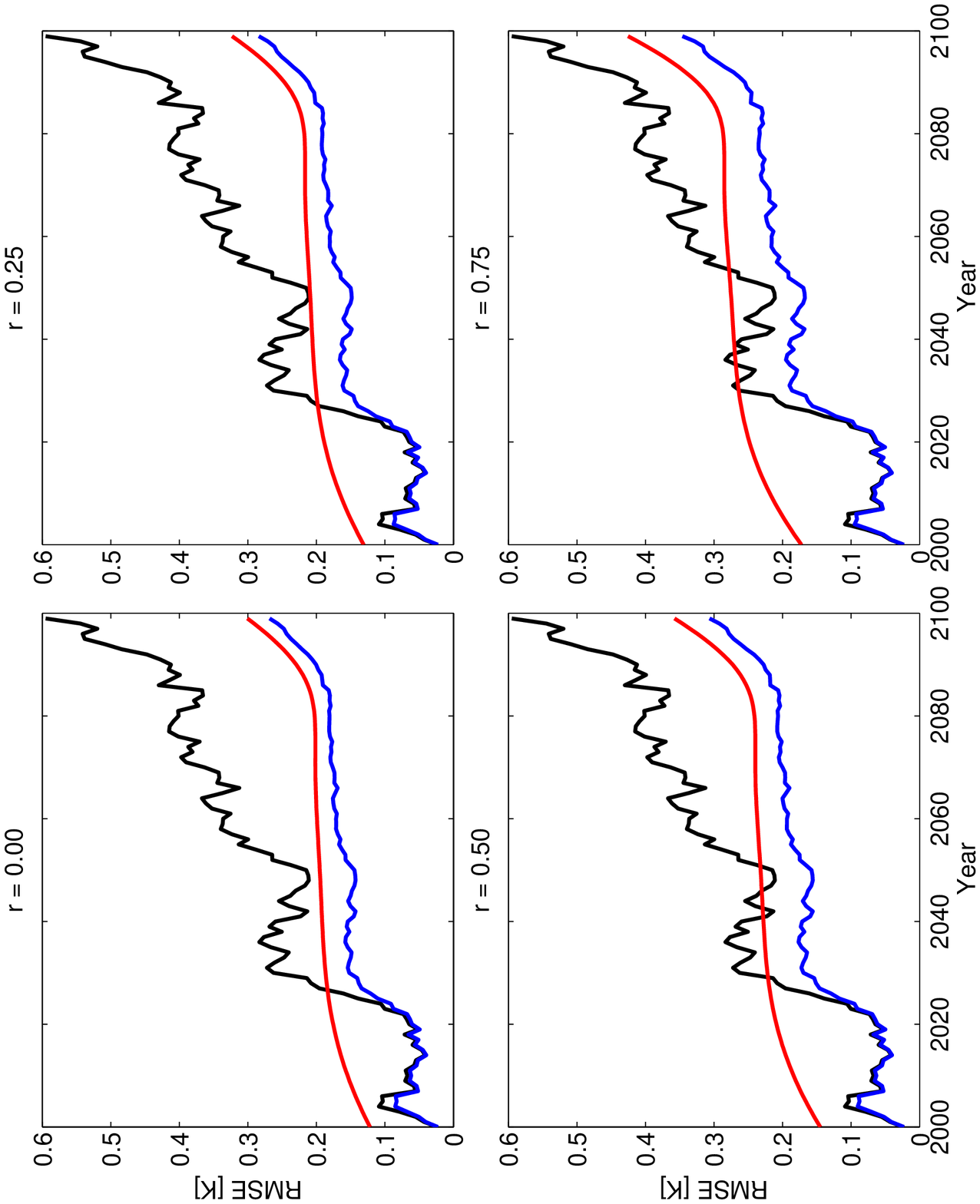}}
\end{center}\caption{
The estimated RMSE of temperature predictions made from the `no change' forecast (black line), the IPCC
ensemble mean (red line) and the damped IPCC ensemble mean (blue line).
}
\label{fig8}\end{figure}

We now repeat the example for winter (DJF) decadal-mean precipitation data for the UK.
Figures~\ref{fig5}, \ref{fig6}, \ref{fig7} and~\ref{fig8} correspond
to figures~\ref{fig1}, \ref{fig2}, \ref{fig3} and~\ref{fig4}, but for precipitation.
In this case $n=14$.

Figure~\ref{fig5} shows a complex temporal structure in the ensemble mean, especially in the early years of the prediction.
This pattern may well not be realistic.
It leads to values of $k$ (see figure~\ref{fig6}) which show an initial peak, then a drop to low values, and
then a rise to roughly constant values from 2030 onwards.
Interpreting these numbers is somewhat difficult, and this detail should perhaps be smoothed in some way.
One approach would be to assume that the temporal structure in the changes in the ensemble mean is not realistic, while the overall
trend is, replace the whole trend in the ensemble mean by a fitted linear trend, and recalculate $k$ values from that.
Another approach would be to avoid such ad-hoc decisions, and use the results as is, and that is what we do here.

The values of $k$ for longer lead-times are relatively stable. They are lower than the corresponding values for
temperature shown in figure~\ref{fig2}. As a result the damped predictions, shown in figure~\ref{fig7}, are damped
more towards `no change'. At short lead times the predictions are reduced to very close to zero under all four scenarios.
At long lead times, for the $r=0.75$ correlation scenario, the predicted change is roughly
divided by two.

The estimated prediction errors shown in figure~\ref{fig8} are very interesting. 
Considering panel (d), we see that 
up to around 2050 (and smoothing the black curve a little by eye), 
the `no change' forecast (the black line) actually beats the ensemble mean (the red line).
This is not because the ensemble mean contains no information, but just because it is so poorly estimated that the estimation error
trumps what information it might contain. This is exactly the situation in which damping is likely to be beneficial. 
Comparing the ensemble mean (the red line) and the damped ensemble mean (the blue line), 
the damped ensemble mean has materially lower errors at all lead times, and particularly so
at short lead times, where the errors are between a factor of two and four smaller.

\section{Discussion}

We have considered the question of how to deal with forecasts from numerical climate models that have significant
uncertainty. This arises, for instance, when a multi-model ensemble shows significant differences between
forecasts from different models. Such cases clearly devalue the ensemble mean as a trustworthy prediction.

We encode this idea mathematically using the idea of a bias-variance tradeoff. Such schemes are widely used
in the statistical predictions for climate used in industry (see, for example, \citet{j72}), but have not, to our knowledge,
been applied to numerical climate models. The scheme can also be viewed in other ways too, such as the idea of making
optimal combinations of simple statistical and numerical model forecasts.

When applied to CMIP3-derived predictions for temperature and precipitation in the UK the method makes 
little difference to the temperature predictions, except in the first two decades, 
but a large and material difference to the precipitation predictions at all lead times.
It very significantly reduces predictions of changes in rainfall at short lead times, and even at long lead-times
reduces predictions of the change in rainfall by a factor of two. The estimated errors of the damped predictions
are much lower than the estimated errors for the ensemble mean prediction.
This is a strong suggestion that CMIP3 ensemble mean precipitation
predictions for the UK are, as they stand, too uncertain to be used, and need to damped back towards `no change' in
order to improve their accuracy.

There are various directions in which this work could be taken. In the current scheme, the numerical model prediction
of a change in climate is damped towards a `simple model' of zero change. An alternative would be to damp towards
a more sophisticated simple model prediction: for temperature that might consist of a linear trend. For precipitation
it is rather hard to detect any change in the observed data (at least in the UK), and so a model of zero change may be the only alternative.
When damping towards a linear trend the amount of damping would then depend on both the
size of the estimated uncertainty on the numerical model predicted change, and the estimated uncertainty
on the change predicted from the statistical trend. There are some interesting questions here: would it make sense to
damp towards \emph{both} a prediction of no change and a linear trend?
Or should the prediction of no change and the linear trend model be combined into a single simple model prediction, using
a prior stage of damping?

Another direction would be applying the ideas presented here to seasonal and decadal forecasts.
In seasonal forecasts it might make sense to damp forecasts back to climatology.
In initialised decadal forecasts it might make sense to damp forecasts back to a best estimate of the forced response.

Finally we note that there has recently been a proposal to adjust future climate forecasts using a different principle~\citep{palmer08}.
It be may possible to relate or combine the two approaches.

\clearpage
\bibliography{IPCC_damping}

\appendix

\clearpage
\section{MSE estimates for the damped forecast}

As a final note, we derive the estimated MSE for the damped forecast.

Substituting
\begin{eqnarray}
k&=&\frac{d^2}{d^2+V(\hat{d})}\\
1-k&=&\frac{V}{d^2+V(\hat{d})}
\end{eqnarray}

into
\begin{equation}
\mbox{MSE}=V(\hat{c})+d^2(k-1)^2+k^2V(\hat{d})+\sigma^2
\end{equation}

gives:

\begin{eqnarray}
\mbox{MSE}
&=&V(\hat{c})+d^2(k-1)^2+k^2V(\hat{d})+\sigma^2\\
&=&V(\hat{c})+d^2\frac{V^2}{(d^2+V(\hat{d}))^2}+\frac{d^4}{(d^2+V(\hat{d}))^2}V(\hat{d})+\sigma^2\\
&=&V(\hat{c})+\frac{Vd^2}{V+d^2}+\sigma^2\\
&=&V(\hat{c})+V \frac{d^2}{V+d^2}+\sigma^2\\
&=&V(\hat{c})+kV+\sigma^2
\end{eqnarray}

This is now the estimated MSE on the damped forecast, and is also the
prediction for the variance that should be used in the damped probabilistic forecast.

\clearpage

\end{document}